\documentclass{article}
\usepackage{spconf,amsmath,epsfig}
\usepackage{multirow}
\begin{document}\sloppy

\def\x{{\mathbf x}}
\def\L{{\cal L}}

\title{MULTI-TARGET EMOTIONAL VOICE CONVERSION WITH NEURAL VOCODERS}
%
\name{\begin{tabular}{c}Songxiang Liu, Yuewen Cao, Helen Meng\end{tabular}}
\address{
   Human-Computer Communications Laboratory \\
   Department of Systems Engineering and Engineering Management,\\
  The Chinese University of Hong Kong, Shatin, N.T., Hong Kong SAR, China \\
  \small \tt \{sxliu, ywcao, hmmeng\}@se.cuhk.edu.hk
  }

\maketitle

\begin{abstract}
Emotional voice conversion (EVC) is one way to generate expressive synthetic speech. Previous approaches mainly focused on modeling one-to-one mapping, i.e., conversion from one emotional state to another emotional state, with Mel-cepstral vocoders. In this paper, we investigate building a multi-target EVC (MTEVC) architecture, which combines a deep bidirectional long-short term memory (DBLSTM)-based conversion model and a neural vocoder. Phonetic posteriorgrams (PPGs) containing rich linguistic information are incorporated into the conversion model as auxiliary input features, which boost the conversion performance. To leverage the advantages of the newly emerged neural vocoders, we investigate the conditional WaveNet and flow-based WaveNet (FloWaveNet) as speech generators. The vocoders take in additional speaker information and emotion information as auxiliary features and are trained with a multi-speaker and multi-emotion speech corpus. Objective metrics and subjective evaluation of the experimental results verify the efficacy of the proposed MTEVC architecture for EVC.
\end{abstract}

\section{Introduction}
\label{sec:intro}

Synthetic speech has burgeoned at a rapid rate for human-computer interaction in recent years. Human speech is a complex signal that contains rich information, which includes linguistic information, para- and non-linguistic information. Linguistic information is explicitly represented by the written language or uniquely inferred from context; para-linguistic information is added by the speaker to modify or supplement the linguistic information, and non-linguistic information is not generally controlled by the speaker, such as the speaker's emotion \cite{fujisaki2004information}. Natural sounding synthetic speech should encompass all these factors. In text-to-speech (TTS) synthesis, there have been many attempts to generate stylish, expressive or emotional speech \cite{Nose2007style, henter2017principles, lorenzo2018investigating, wang2018style, skerry2018towards, wu2018feature, Wu2018}. Another way to synthesize natural sounding speech is adopting emotional voice conversion (EVC) techniques, which aims at converting speech from one emotional state into another one, keeping the basic linguistic information and speaker identity.

There has been tremendous active research in EVC. Since prosody plays an important role in conveying various types of non-linguistic information, typical EVC approaches focus on modeling the conversion of short-time spectral features and prosodic features, such as the F0 contour and energy contour, jointly or disjointly.
In \cite{tao2006prosody}, a Gaussian mixture model (GMM) and a classification regression tree model were adopted to model the F0 contour conversion from neutral speech to emotional speech. 
In \cite{inanoglu2007system}, the F0 contour was modeled and generated by context-sensitive syllable-based HMMs, the duration was transformed using phone-based relative decision trees, and the spectrum was converted using a GMM-based or a codebook selection approach. 
Prosody is inherently supra-segmental and hierarchical in nature, of which the conversion is affected by both short- and long-term dependencies.
There have been many attempts to model prosody in multiple temporal levels, such as the phone, syllable and phrase levels \cite{latorre2008multilevel, wu2010hierarchical, obin2011stylization, qian2011improved}. 
Continuous wavelet transform (CWT) can effectively model F0 and energy contour in different temporal scales. CWT was adopted for F0 modeling within the non-negative matrix factorization (NMF) model \cite{ming2015fundamental}, and for F0 and energy contour modeling within a deep bidirectional LSTM (DBLSTM) model \cite{ming2016deep}. 
Using CWT method to decompose the F0 in different scales has also been explored in \cite{luo2017emotional, luo2016emotional}, where neural networks (NNs) or deep belief networks (DBNs) were adopted.

While previous approaches have shown their effectiveness for EVC, one limitation is that they only model one-to-one mapping, i.e., conversion from one emotional state to another emotional state. An individual model is needed for each emotional conversion pair, where a sufficient number of parallel training samples of the studied emotional states are needed to get desirable conversion performance. Possible way to tackle this limitation is to train a single multi-target EVC model using multi-emotion speech corpus. Another limitation of previous approaches is that they adopt Mel-cepstrum vocoders, such as STRAIGHT \cite{kawahara1999restructuring} and WORLD \cite{morise2016world}, to compute acoustic features (spectrum, F0, etc.) and synthesize converted speech waveform. Such vocoders impose many assumptions based on prior knowledge specific to speech, e.g., fixed-length analysis window, time-invariant linear filters, stationary Gaussian process, etc. \cite{tamamori2017speaker}. Moreover, phase information of the original speech is lost from the extracted acoustic features, which degrades the naturalness of the synthesized speech. Replacing the Mel-cepstrum vocoders with neural vocoders, such as WaveNet \cite{van2016wavenet}, may be helpful. Current state-of-the-art TTS architectures commonly use the WaveNet vocoder with a Mel-spectrogram as conditioning for high-fidelity speech synthesis \cite{shen2018natural}. Although Mel-spectrograms discard phase information, previous study \cite{tamamori2017speaker} shows that WaveNet conditioned on acoustic features is capable of recovering phase information and generates more natural speech compared to Mel-cepstrum vocoders. Since WaveNet requires an autoregressive sampling scheme, it can not fulfill the demand of real-time waveform generation. The very recent flow-based neural vocoder \cite{prenger2018waveglow, kim2018flowavenet} is a possible alternative, which requires only a single maximum likelihood loss for training and can efficiently sample raw audio in real-time.

In this paper, we investigate building a multi-target emotional VC (MTEVC) architecture, which combines a deep bidirectional long-short term memory (DBLSTM)-based conversion model and a neural vocoder. Phonetic posteriorgrams (PPGs) obtained from a speaker-independent automatic speech recognition (SI-ASR) system are speaker-independent linguistic features, which have been successfully used for VC with non-parallel data\cite{sun2016phonetic, liu2018voice, liu2018hccl, liu2019jointly}.
The proposed conversion model takes in source Mel-spectrograms and their corresponding PPGs as inputs, and outputs the predicted target Mel-spectrograms. To control output emotional states, the conversion model takes in emotion codes as additional auxiliary features in the form of one-hot representations.
We investigate two kinds of neural vocoders for EVC in this study, which are WaveNet and flow-based WaveNet (FloWaveNet) \cite{kim2018flowavenet}. Mel-spectrograms are used as conditions. To get high-fidelity generated speech, sufficient amounts of speech data are needed for training. Since it is often difficult to collect considerable amount of speech data of one particular emotion state from a speaker, we propose to train the neural vocoders with multi-speaker and multi-emotion speech data, where speaker and emotion codes are injected as auxiliary features.
It is expected that the use of speaker codes and emotion codes enables the neural vocoders to capture speaker-dependent and emotion-dependent temporal structures such as phase information, which is not represented in Mel-spectrograms. 
The contributions of this paper are in the incorporation of two kinds of auxiliary speaker and emotion codes into neural vocoders to achieve multi-speaker and multi-emotion training and as such achieve multi-target EVC.

The rest of the paper is organized as follows: Section 2 introduces the proposed MTEVC architecture. Section 3 describes the implementation details. Section 4 gives the experimental results and analysis. Section 5 draws the conclusions.

\begin{figure}[t]
  \centering
  \centerline{\includegraphics[width=8cm]{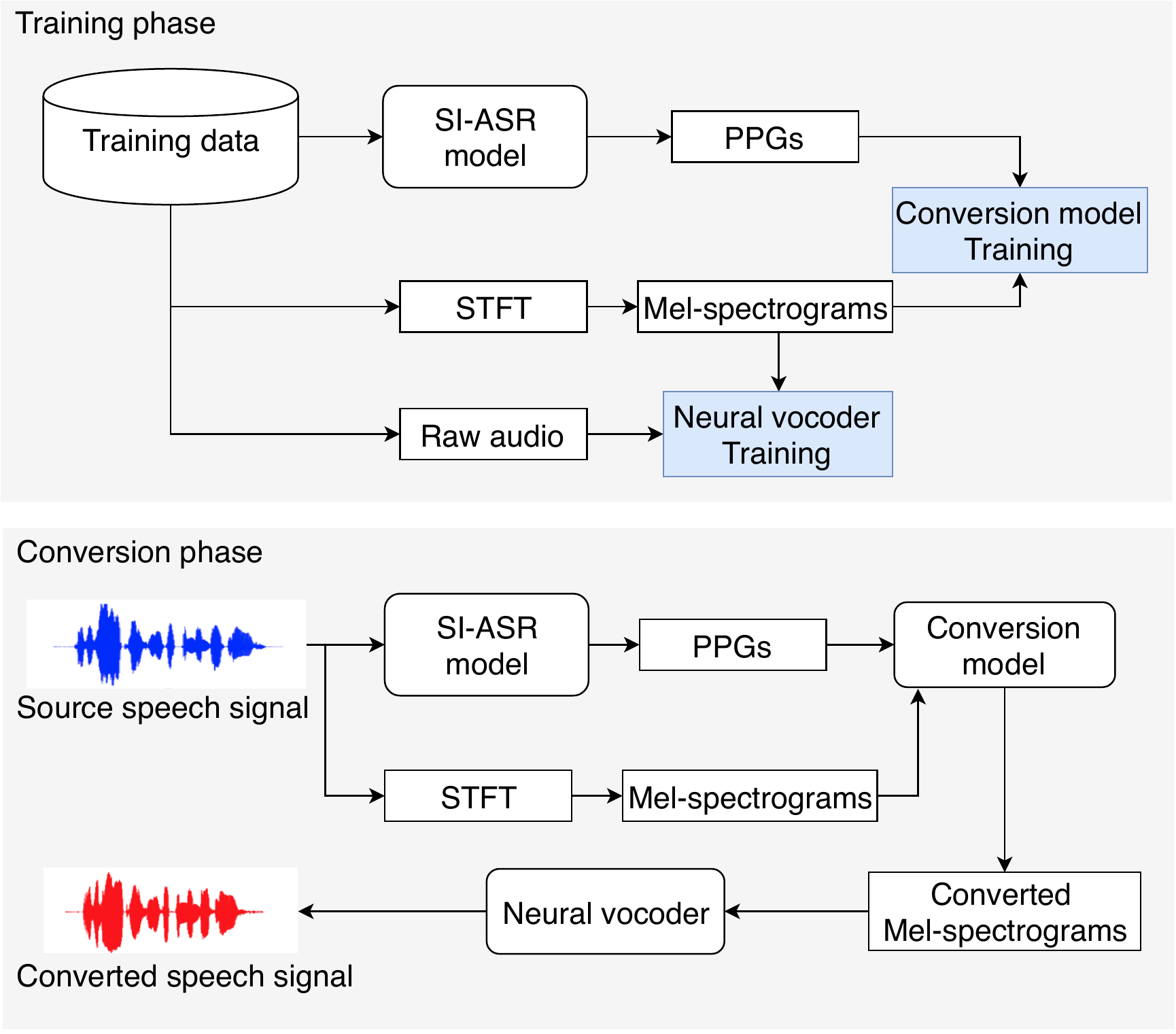}}
  \caption{The training phase and conversion phase of the proposed multi-target emotional voice conversion architecture.}
  \label{fig:sysarch}
\end{figure}


\section{MULTI-TARGET EMOTIONAL VOICE CONVERSION ARCHITECTURE}
\subsection{Overview of architecture}
The proposed MTEVC system consists of two phases: training phase and conversion phase, as shown in Fig.~\ref{fig:sysarch}. During the training phase, three models are trained, which are SI-ASR model, conversion model and neural vocoder. The SI-ASR model is adopted to extract PPGs from the speech data, which is trained with a standard ASR corpus.
At conversion phase, the source speech signal drives the SI-ASR model to obtain its PPG representation. The source Mel-spectrograms and the obtained PPGs are then fed into the trained conversion model to predict the converted Mel-spectrograms. The converted speech signal is obtained using the trained neural vocoders with the converted Mel-spectrograms as auxiliary features. 

\subsection{The Conversion model}

\begin{figure}[t]
  \centering
  \centerline{\includegraphics[width=8.3cm]{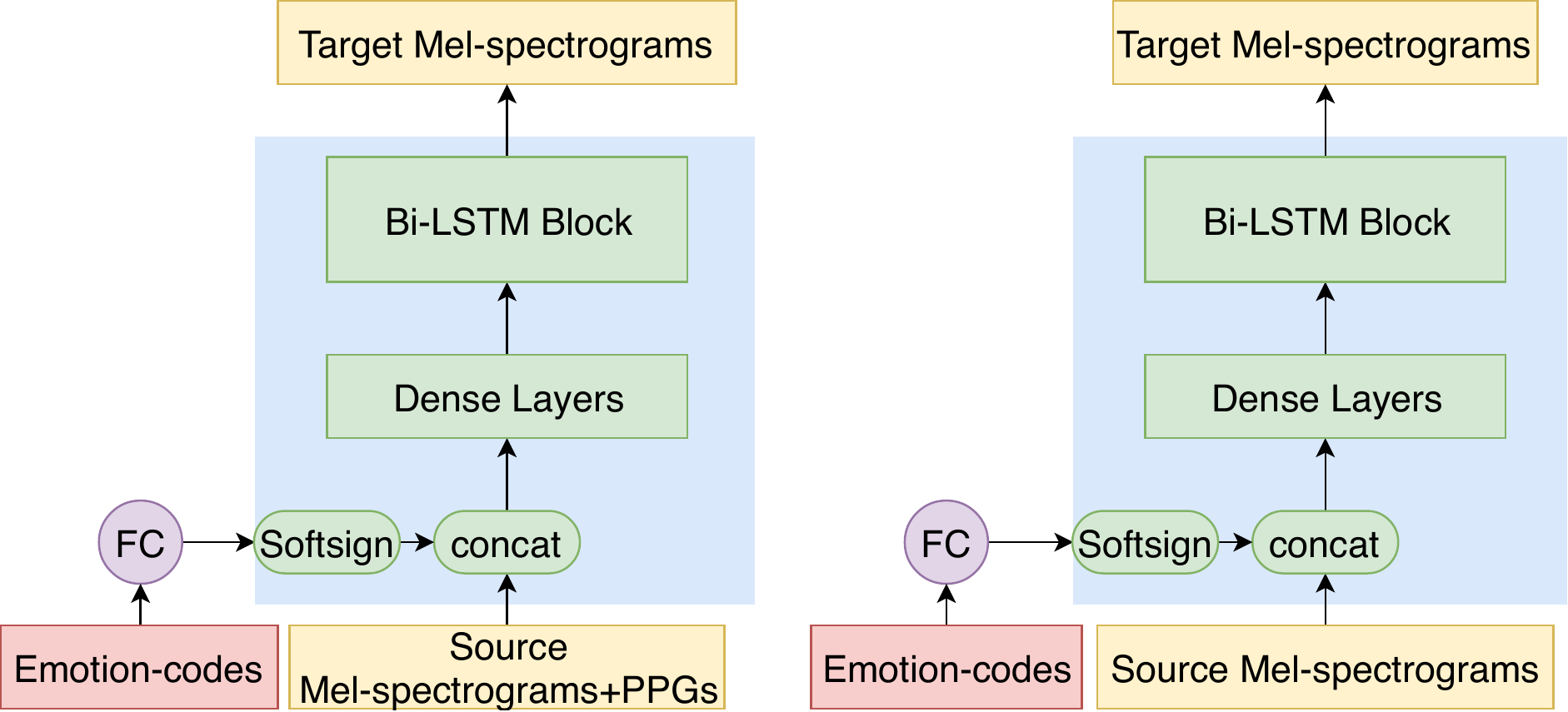}}
  \caption{Left: the proposed conversion model. Right: the baseline conversion model.}
  \label{fig:evcmdl}
\end{figure}

The only difference between the baseline system and the proposed system lies in the conversion models, which are shown in Fig.~\ref{fig:evcmdl}.
The proposed conversion model takes Mel-spectrograms and the corresponding PPGs extracted from the source emotional speech as inputs and Mel-spectrograms computed from the target emotional utterance as outputs. The inputs of the baseline conversion model take only source Mel-spectrograms. The bottleneck features extracted by an ASR model contain high-level and linguistic-related information, which have been shown to help achieve stable VC results \cite{zhang2018sequence}. Since rich linguistic information are contained in the PPGs, it is expected that the conversion model can achieve better conversion performance with PPGs as auxiliary features. The conversion model takes one-hot represented emotion code corresponding to the target emotion as additional auxiliary input feature. Since the number of time frames of the source and target Mel-spectrograms are different, a time alignment approach, such as dynamic time warping (DTW), is required to align the source-target features before training the conversion model.

The model architecture contains several dense layers and bidirectional LSTM layers, which has similar architecture configuration to that in \cite{liu2018voice}, as illustrated in Fig.~\ref{fig:evcmdl}. The emotion codes are first fed into the emotion embedding layer to get the real-valued vector representations, and then transformed by a fully connected layer with softsign activation function. The obtained emotion conditionings are concatenated with the input Mel-spectrograms and PPGs before being fed into the dense layers. The model parameters are updated by minimizing L1 loss between the predicted Mel-spectrograms and the target ground-truth Mel-spectrograms using back-propagation.

\begin{figure}[t]
  \centering
  \centerline{\includegraphics[width=5.5cm]{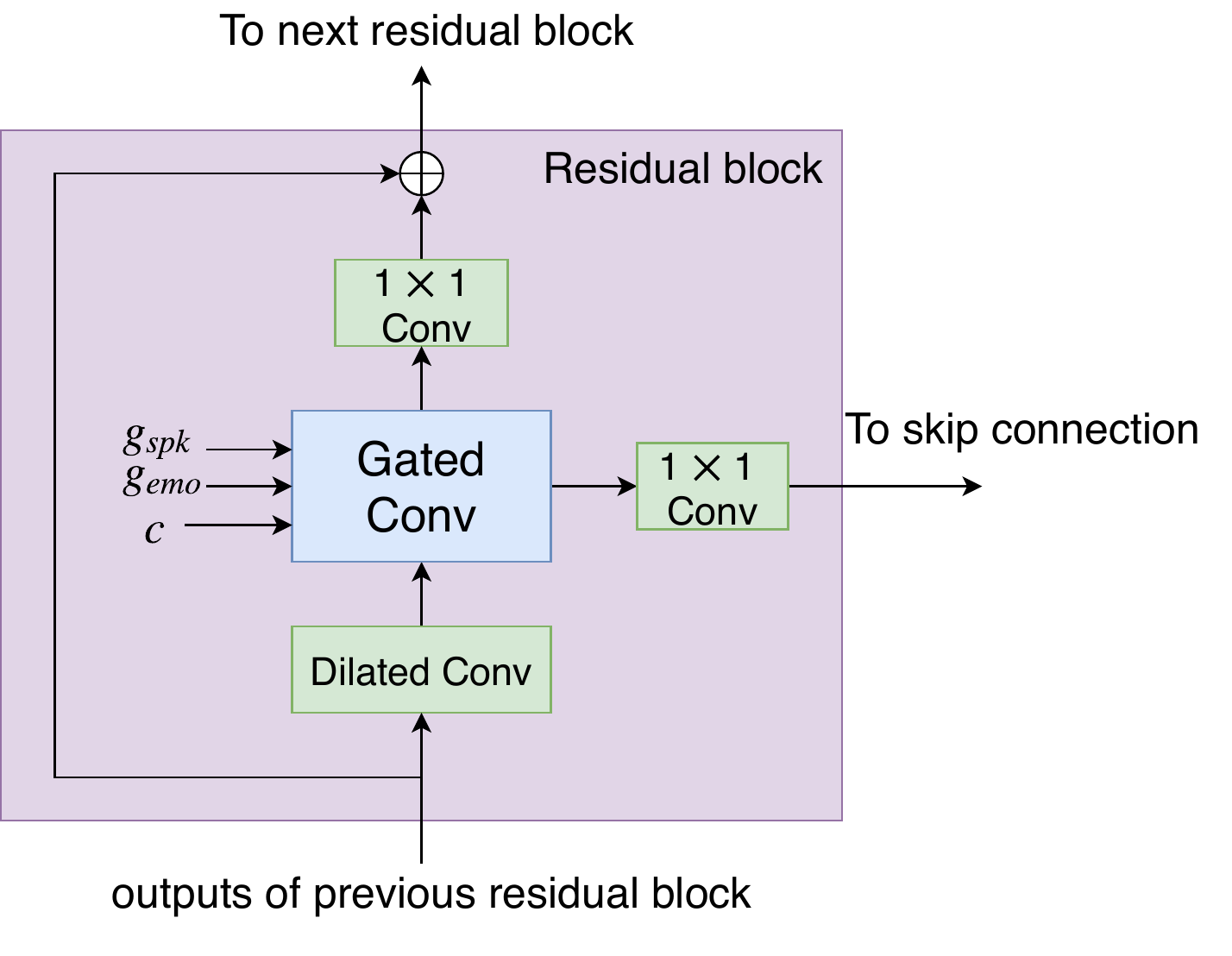}}
  \caption{The residual block of the conditional WaveNet vocoder and the conditional FloWaveNet vocoder, where $\mathbf{c}$ denotes the local Mel-spectrograms conditioning, $g_{spk}$ denotes the global speaker conditioning and $g_{emo}$ denotes the global emotion conditioning, respectively.}
  \label{fig:resblock}
\end{figure}

\subsection{Neural vocoders}
We investigate two kinds of neural vocoders: the WaveNet vocoder and the recently proposed real-time flow-based WaveNet (FloWaveNet) vocoder \cite{prenger2018waveglow, kim2018flowavenet}. These neural vocoders are generative models directly modeling raw audio samples. By conditioning on auxiliary features, the characteristics of generated samples can be controlled. In this paper, we take Mel-spectrograms as local conditionings, and speaker codes and emotion codes as global conditionings, respectively. 
Both of the WaveNet vocoder and the FloWaveNet vocoder have residual block as illustrated in Fig.~\ref{fig:resblock}, where the local Mel-spectrograms $\mathbf{c}$, global speaker conditioning $g_{spk}$ and global emotion conditioning $g_{emo}$ are element-wisely added to the output of the dilated convolution layers before being fed into the gated convolution layers. In the next subsections, we first introduce the conditional WaveNet vocoder, and then introduce the conditional FloWaveNet vocoder. 


\subsubsection{Conditional WaveNet}
Given a waveform $\mathbf{x} = \{x_1,x_2, ..., x_T\}$, WaveNet models the distribution of $\mathbf{x}$ conditioned on the history audio samples and additional auxiliary features $\mathbf{h}$ as
\begin{equation}
  p(\mathbf{x}|\mathbf{h}) = \prod_{t=1}^{T}p(x_t|x_1, x_2, ..., x_{t-1}, \mathbf{h}).
\end{equation}
The auxiliary features $\mathbf{h}$ in this paper are Mel-spectrograms, speaker codes and emotion codes. 
The conditionings are incorporated to the gated convolution layers, whose outputs
\begin{equation}
  \mathbf{z} = \tanh(W_{f,k}*\mathbf{x}+V_{f,k}*\mathbf{y}) \odot \sigma(W_{g,k}*\mathbf{x}+V_{g,k}*\mathbf{y}),
\end{equation}
where $W$ and $V$ are trainable convolution filters, $V_{f,k}*\mathbf{y}$ represent $1\times1$ convolution, and $\mathbf{y}$ represents transformed auxiliary features which have same temporal resolution as the input speech waveform with the function $\mathbf{y} = f(\mathbf{h})$.

\subsubsection{Conditional FloWaveNet}

\begin{figure}[t]
  \centering
  \centerline{\includegraphics[width=8cm]{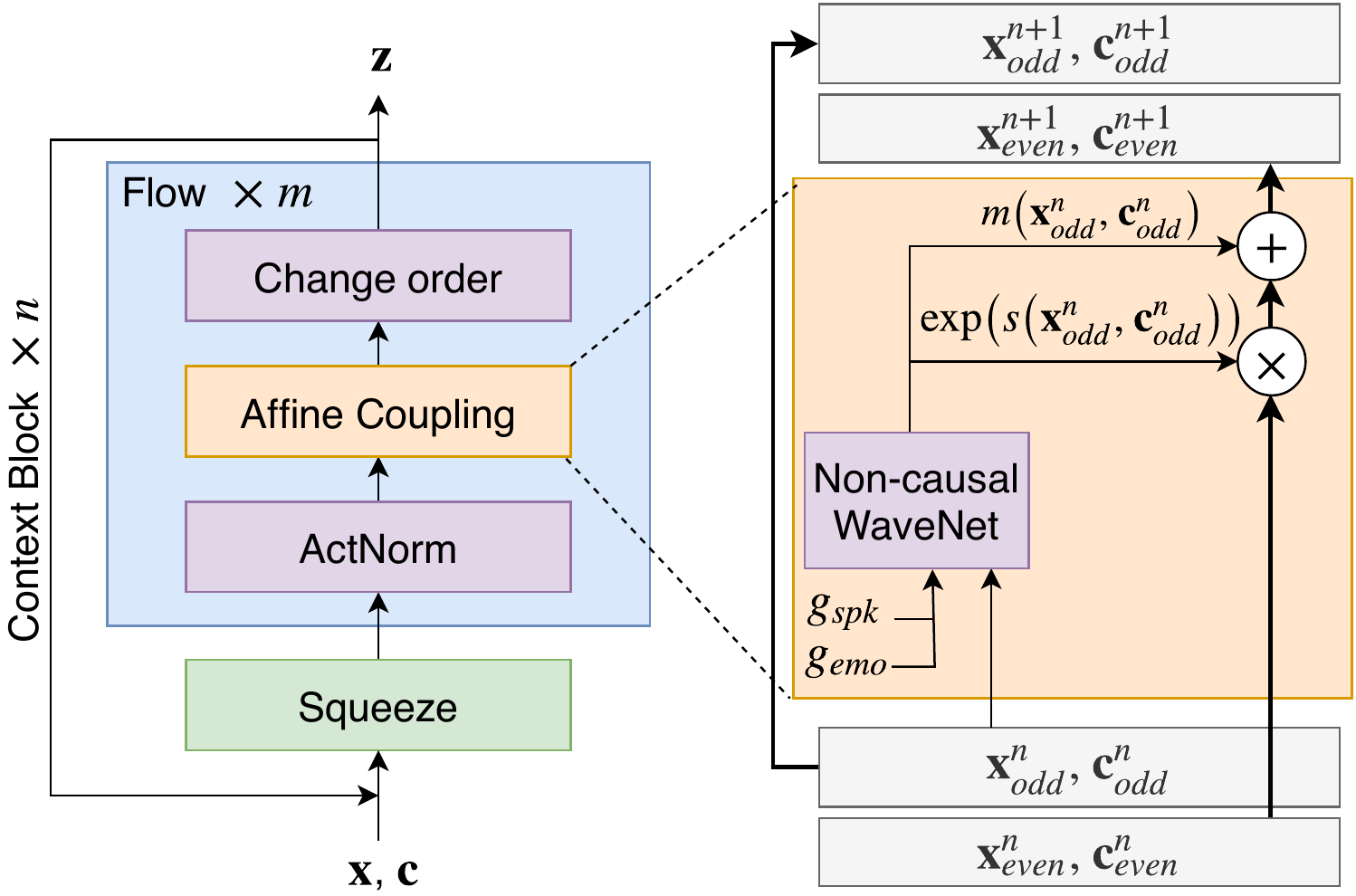}}
  \caption{Schematic diagram of the conditional FloWaveNet vocoder, where $\mathbf{x}$ denotes raw audio signal, $\mathbf{c}$ denotes the local Mel-spectrograms conditioning, $g_{spk}$ denotes the global speaker conditioning and $g_{emo}$ denotes the global emotion conditioning, respectively.}
  \label{fig:flowavenet}
\end{figure}

In this paper, we follow the basic FloWaveNet architecture proposed in \cite{kim2018flowavenet}. The FloWaveNet is a hierarchical architecture composed of $n$ context blocks, each of which has one squeeze operation and $m$ invertible flows, as illustrated in Fig.~\ref{fig:flowavenet}. The squeeze operation doubles the channel dimension of the audio data and conditions by splitting the time dimension in half. Given a waveform audio signal $\mathbf{x}$, assume there is an invertible transformation function $f(\mathbf{x})$: $\mathbf{x} \rightarrow \mathbf{z}$, where the prior distribution of $\mathbf{z}$ is known and its likelihood can be calculated conveniently, e.g., $\mathbf{z} \sim \mathcal{N}(0, \bf{I})$. By change of variables, the log likelihood of $\mathbf{x}$ has the following form:
\begin{equation}
  \log p_{\theta}(\mathbf{x}) = \log p_{\theta}(\mathbf{z}) + \log|\det(\mathbf{J}(f(\mathbf{x})))|,
\end{equation}
where $\mathbf{J}$ is the Jacobian. To make the calculation of the Jacobian and the inverse transform $\mathbf{x}=f^{-1}(\mathbf{z})$ tractable, the affine coupling layer in each flow is novelly designed. 

Specifically, each flow operation contains an activation normalization (ActNorm) \cite{kingma2018glow}, affine coupling layer \cite{dinh2016density}, and change order operation. The ActNorm layer is a per-channel parametric affine transformation stabilizing training. The affine coupling layer, which is designed to be bijective, includes a non-causal WaveNet module as shown in Fig.~\ref{fig:flowavenet}. The local Mel-spectrograms conditionings, as well as the global speaker and emotion conditionings are added to the residual blocks in the way mentioned above. Each layer is the parametric transform $f_n:\mathbf{x}^n \rightarrow \mathbf{x}^{n+1}$, which keeps a half of the channel dimension identical and applies an affine transform only on the remaining half, as the following:
\begin{equation}
  \mathbf{x}_{odd}^{n+1} = \mathbf{x}_{odd}^{n},
\end{equation}
\begin{equation}
  \mathbf{x}_{even}^{n+1} = \mathbf{x}_{even}^{n}\odot \exp(s(\mathbf{x}_{odd}^{n}, \mathbf{c}^{n})) + m(\mathbf{x}_{odd}^{n}, \mathbf{c}^{n}),
\end{equation}
where $m$ and $s$ stand for translation and scale, and $\odot$ is element-wise product.
The inverse transform $f_{n}^{-1}:\mathbf{x}^{n+1}\rightarrow \mathbf{x}^{n}$ has the following nice form:
\begin{equation}
  \mathbf{x}_{odd}^{n} = \mathbf{x}_{odd}^{n+1},
\end{equation}
\begin{equation}
  \mathbf{x}_{even}^{n} = \frac{\mathbf{x}_{even}^{n+1} - m(\mathbf{x}_{odd}^{n+1}, \mathbf{c}_{odd}^{n+1})}{\exp(s(\mathbf{x}_{odd}^{n+1}, \mathbf{c}^{n+1}))}.
\end{equation}
The Jacobian matrix is lower triangular and the determinant is a product of the diagonal elements. The change order operation swaps the order of $\mathbf{x}_{odd}$ and $\mathbf{x}_{even}$ so that all channels can affect each other during the subsequent flow operations. The FloWaveNet model requires only a single maximum likelihood loss without any additional auxiliary terms for training and is inherently parallel due to the flow-based transformation during generation time.

\section{IMPLEMENTATION}

The SI-ASR model has a DNN architecture with 4 hidden layers containing 1024 hidden units. Tied tri-phone states are treated as the phonetic class of PPGs. The dimension of PPGs is 131. Speech data having sampling rate 16kHz from 462 speakers in the TIMIT corpus \cite{garofolotimit} is used for training. Mel-frequency cepstral coefﬁcients (MFCCs) of dimension 13 and their first and second derivatives are used as features. We use a 25-ms Hamming window with 5-ms frame shift.

We use the CASIA Chinese Emotional Corpus, recorded by the Institute of Automation, Chinese Academy of Sciences, where each sentence with the same semantic texts is spoken by 2 female and 2 male speakers in six different emotional tones: happy, sad, angry, surprise, fear, and neutral. The sampling rate is 16kHz. We use 260 utterances from one female speaker for each emotion to train the conversion models, 20 utterances as validation set and another 20 utterances as evaluation set.
The conversion models have 2 dense layers with 256 hidden units and $\tanh$ activation function. The Bi-LSTM block has 4 layers, having 256 hidden units in each direction. The emotion embedding dimension is 16. The conversion model is trained with Adam optimizer with learning rate 0.001.

We use 360 utterances for each speaker and each emotion state from the CASIA corpus to train the conditional WaveNet vocoder and the conditional FloWaveNet vocoder. We set both the speaker embedding dimension and the emotion embedding dimension to be 16. The local Mel-spectrogram conditionings, which have dimension of 80, are adjusted to have the same time resolution as the audio signals by transposed convolutions. The WaveNet has a 24-layer architecture with four 6-layer dilation cycles. The 8-bit $\mu$-law quantized waveform is used for WaveNet training. The FloWaveNet has 8 context blocks, each of which has 6 flows. The non-causal WaveNet module in the affine coupling layers has 2-stack architecture with a kernel size of 3. We set the residual, skip, and gate channels to be 256 in the WaveNet architecture. Adam optimizer with initial learning rate of 0.001 is used for training the WaveNet and the FloWaveNet. We schedule the learning rate decay by a factor of 0.5 for every 100K steps.

\section{Experimental results and analysis}
Previous study \cite{liu2018hccl} has shown that the Griffin-Lim algorithm \cite{griffin1984signal} achieves better VC performance than the Mel-cepstral vocoder STRAIGHT. Therefore, we compare the baseline and the proposed conversion models, using the Griffin-Lim algorithm, WaveNet and FloWaveNet vocoders as different waveform generation techniques.

\subsection{Objective evaluation}
The Mel Cepstral Distortion (MCD) is used for the objective evaluation of spectral conversion. The MCD is computed as:
\begin{equation}
  MCD  =  \frac{10}{\ln 10} \sqrt{2\sum_{i=1}^{M}(\mathbf{MCC}_i^t - \mathbf{MCC}_i^c)^2},
\end{equation}
where $\mathbf{MCC}_i^t$ and $\mathbf{MCC}_i^c$ represent the target and the converted Mel-cepstral computed from the target recordings and converted waveform, respectively.
The LogF0 mean squared error (MSE) is computed to evaluate the F0 conversion, which has the form 
\begin{equation}
  MSE = \frac{1}{N} \sum_{i=1}^{N}(\log(F0_i^t) - \log(F0_i^c))^2,
\end{equation}
where $F0_i^t$ and $F0_i^c$ denote the target and the converted F0 features, respectively. The average MCD and LogF0-MSE results are illustrated in Table \ref{table:1}, where neutral speech is converted to five different emotion states, i.e., sad, angry, surprise, fear and happy.

\begin{table}[t]
  \caption{\label{table:1} {MCD and LogF0-MSE results. B denotes the baseline conversion model while P denotes the proposed conversion model. GL means using Griffin-Lim algorithm to generate converted waveform.}}
  \center
  \small
\begin{tabular}{|c|c|c|c|c|c|}
\hline  
\multirow{2}{*}{} & \multicolumn{5}{c|}{MCD (dB)} \\ 
                  \cline{2-6}
                  & Sad          & Angry        & Surprise    & Fear      & Happy         \\
                  \hline
B-GL              & 10.71        & 11.23        & 11.58       & 10.66     & 11.32        \\
B-WaveNet         & 8.94         & 9.15         & 9.42        & 8.89      & 9.32         \\
B-FloWaveNet      & 8.09         & 8.78         & 8.84        & 8.11      & 8.63         \\
P-GL              & 10.70        & 10.78        & 11.15       & 10.56     & 10.87        \\
P-WaveNet         & 8.85         & 9.32         & 9.18        & 8.84      & 9.18         \\
P-FloWaveNet      &\bf 8.05      & \bf 8.66     & \bf 8.78    & \bf 8.02  & \bf 8.62     \\
\hline

\hline  
\multirow{2}{*}{} & \multicolumn{5}{c|}{LogF0-MSE} \\ 
                  \cline{2-6}
                  & Sad          & Angry        & Surprise    & fear      & Happy         \\
                  \hline
B-GL              & 1.66         & 0.99         & 0.98        & 1.74      & 1.05        \\
B-WaveNet         & 1.11         & 0.98         & 0.87        & 1.12      & 0.91         \\
B-FloWaveNet      & 1.23         & 2.79         & 2.63        & 1.09      & 1.51         \\
P-GL              & 1.85         & \bf 0.87     & \bf 0.71    & 1.72      & 0.97        \\
P-WaveNet         & 1.11         & 0.89         & 1.03        & 1.04      & \bf 0.78     \\
P-FloWaveNet      & \bf 1.05     & 2.74         & 3.23        & \bf 0.87  & 1.96         \\
\hline
\end{tabular}
\end{table}

\begin{table}[t]
  \caption{\label{table:2} {Subjective classification results}}
  \small
   \center
\begin{tabular}{|c|c|c|c|c|}
\hline 
\multicolumn{2}{|c|}{Target \textbackslash Perception} & Happy & Sad & Neutral \\
\hline
\multirow{2}{*}{B-GL}   & Happy      & {\bf{76.8}}\%     & 15.8\%        & 7.4\%         \\
                            \cline{2-5}
                            & Sad    & 23.3\%      & {\bf{65.8}}\%       & 10.9\%         \\
                            \hline
\multirow{2}{*}{B-WaveNet}   & Happy      & {\bf{76.7}}\%     & 11.1\%   & 12.2\%         \\
                            \cline{2-5}
                            & Sad    & 10.0\%      & {\bf{51.1}}\%       & 38.9\%         \\
                            \hline
\multirow{2}{*}{B-FloWaveNet} & Happy      & {\bf{78.9}}\%     & 6.7\%        & 14.4\%         \\
                            \cline{2-5}
                            & Sad    & 8.8\%      & {\bf{75.6}}\%       & 15.6\%         \\
                            \hline
\multirow{2}{*}{P-GL}      & Happy      & {\bf{77.8}}\%     & 1.1\%        & 21.1\%         \\
                            \cline{2-5}
                            & Sad    & 4.4\%      & {\bf{70.0}}\%       & 25.6\%         \\
                            \hline
\multirow{2}{*}{P-WaveNet} & Happy      & {\bf{86.7}}\%     & 3.3\%        & 10.0\%         \\
                            \cline{2-5}
                            & Sad    & 22.2\%     & {\bf{62.2}}\%       & 15.6\%         \\
                            \hline
\multirow{2}{*}{P-FloWaveNet} & Happy      & {\bf{78.9}}\%     & 4.4\%        & 16.7\%         \\
                            \cline{2-5}
                            & Sad    & 9.9\%      & {\bf{80.1}}\%       & 10.0\%         \\
\hline 
\end{tabular}
\end{table}

Based on the MCD results, the proposed conversion model with FloWaveNet as waveform generator (P-FloWaveNet) gets the best performance consistently across all emotion pairs in terms of spectral conversion. The baseline conversion model with Griffin-Lim (B-GL) gets the worst conversion performance. Comparing MCDs of different waveform generation mechanisms in the baseline or the proposed conversion frameworks, we can see that the neural vocoders are expected to achieve better spectral conversion performance than the Griffin-Lim algorithm. This means that the neural vocoders, which are trained with speech data, are capable of recovering the discarded information from the Mel-spectrograms (e.g., phase information) better than the Griffin-Lim algorithm. The recently proposed FloWaveNet model, which outperforms the autoregressive WaveNet model in spectral conversion, is promising and deserves more investigation. We can also see that adding PPGs as auxiliary input features can help boost spectral conversion performance.
Based on the LogF0-MSE results, the best pitch conversion performance of different emotion conversion pairs are achieved by different models, but all by the proposed conversion model. This means that incorporating linguistic information into the conversion model enables better pitch conversion. 
The objective metrics validate the effectiveness of achieving multi-target EVC using the proposed models.

\subsection{Subjective evaluation}
A subjective emotion classification test is conducted, where we choose two conversions (neutral-to-happy and neutral-to-sad) and each model has 20 testing utterances (10 for each conversion). The listeners are asked to label the stimuli as more 'happy' or more 'sad' when compared with a neutral reference. 15 native mandarin Chinese speakers take part in this test. The subjective classification results are shown in Table \ref{table:2}. 

According to the subjective evaluation results, the proposed conversion model with WaveNet (P-WaveNet) achieves the best result for neutral-to-happy conversion, while the P-FloWaveNet model achieves the best result for neutral-to-sad conversion. The B-FloWaveNet model also achieves good evaluation results for neutral-to-happy and neutral-to-sad conversions, with degradation by 8.9\% and 5.6\%, respectively. Comparing P-GL with B-GL, P-WaveNet with B-WaveNet and P-FloWaveNet with B-FloWaveNet, we can see that the proposed conversion models taking PPGs as auxiliary inputs achieve better results under both the Griffin-Lim and the neural waveform generation frameworks. Subjective classification results also validate the efficacy of multi-target EVC using the proposed models.

\section{CONCLUSIONS}

In this paper, we investigate building a multi-target EVC (MTEVC) architecture, which combines a DBLSTM-based conversion model and a neural vocoder. The target emotion is controlled by an emotion code. PPGs are regarded as auxiliary input features of the conversion model. To make use of multi-speaker and multi-emotion data for training, we study the feasibility of training the WaveNet and FloWaveNet models by incorporating two kinds of codes. Objective and subjective evaluations validate the effectiveness of the proposed approaches for MTEVC. The neural vocoder FloWaveNet, achieves the best spectral conversion performance for all emotion conversion pairs according to the objective metrics. Since the FloWaveNet model enables parallel audio signal generation and requires only a maximum likelihood loss for training, it deserves more investigation for better generation quality, which will be our future work.

\bibliographystyle{IEEEbib}
\footnotesize
\bibliography{icme2019}
\end{document}